\documentstyle[12pt]{article}
\def\lsim{\mathrel{\rlap {\raise.5ex\hbox{$ < $}}
{\lower.5ex\hbox{$\sim$}}}}
\def\gsim{\mathrel{\rlap {\raise.5ex\hbox{$ > $}}
{\lower.5ex\hbox{$\sim$}}}}

\def\sqr#1#2{{\vcenter{\vbox{\hrule height.#2pt
        \hbox{\vrule width.#2pt height#1pt \kern#1pt
           \vrule width.#2pt}
        \hrule height.#2pt}}}}

\def\lsim{{\displaystyle
{{\raise-8pt\hbox{$ <$}}
\atop{\raise5pt\hbox{$\sim$}}}}}
\def\gsim{{\displaystyle
{{\raise-8pt\hbox{$ >$}}
\atop{\raise5pt\hbox{$\sim$}}}}}
\def\slsim{{\displaystyle
{{\raise-8pt\hbox{$\scriptstyle <$}}
\atop{\raise5pt\hbox{$\scriptstyle \sim$}}}}}
\def\sgsim{{\displaystyle
{{\raise-8pt\hbox{$\scriptstyle  >$}}
\atop{\raise5pt\hbox{$\scriptstyle \sim$}}}}}
\newcommand{\oao}[2]{{#1\atopwithdelims[]#2}}
\newcommand{\sump}[0]{\sum_{(h,g)}\!{\raise 4pt \hbox{$'$}}\,}
\catcode`\@=11
\newcount\hour
\newcount\minute
\newtoks\amorpm
\hour=\time\divide\hour by60
\minute=\time{\multiply\hour by60 \global\advance\minute by-\hour}
\edef\standardtime{{\ifnum\hour<12 \global\amorpm={am}%
        \else\global\amorpm={pm}\advance\hour by-12 \fi
        \ifnum\hour=0 \hour=12 \fi
        \number\hour:\ifnum\minute<10 0\fi\number\minute\the\amorpm}}
\edef\militarytime{\number\hour:\ifnum\minute<10 0\fi\number\minute}
\def\draftlabel#1{{\@bsphack\if@filesw {\let\thepage\relax
   \xdef\@gtempa{\write\@auxout{\string
      \newlabel{#1}{{\@currentlabel}{\thepage}}}}}\@gtempa
   \if@nobreak \ifvmode\nobreak\fi\fi\fi\@esphack}
        \gdef\@eqnlabel{#1}}
\def\@eqnlabel{}
\def\@vacuum{}
\def\draftmarginnote#1{\marginpar{\raggedright\scriptsize\tt#1}}
\def\draft{\oddsidemargin -.2truein
        \def\@oddfoot{\sl preliminary draft \hfil
        \rm\thepage\hfil\sl\today\quad\militarytime}
        \let\@evenfoot\@oddfoot \overfullrule 3pt
        \let\label=\draftlabel
        \let\marginnote=\draftmarginnote
   \def\@eqnnum{(\theequation)\rlap{\kern\marginparsep\tt\@eqnlabel}%
\global\let\@eqnlabel\@vacuum}  }
\def\preprint{\twocolumn\sloppy\flushbottom\parindent 2em
        \leftmargini 2em\leftmarginv .5em\leftmarginvi .5em
        \oddsidemargin -.5in    \evensidemargin -.5in
        \columnsep .4in \footheight 0pt
        \textwidth 10.in        \topmargin  -.4in
        \headheight 12pt \topskip .4in
        \textheight 6.9in \footskip 0pt
        \def\@oddhead{\thepage\hfil\addtocounter{page}{1}\thepage}
        \let\@evenhead\@oddhead \def\@oddfoot{} \def\@evenfoot{} }

\def\numberbysection{\@addtoreset{equation}{section}
        \def\theequation{\thesection.\arabic{equation}}}

\def\underline#1{\relax\ifmmode\@@underline#1\else
        $\@@underline{\hbox{#1}}$\relax\fi}

\def\titlepage{\@restonecolfalse\if@twocolumn\@restonecoltrue
\onecolumn
     \else \newpage \fi \thispagestyle{empty}\c@page\z@
        \def\thefootnote{\fnsymbol{footnote}} }

\def\endtitlepage{\if@restonecol\twocolumn \else \newpage \fi
        \def\thefootnote{\arabic{footnote}}
        \setcounter{footnote}{0}}  %\c@footnote\z@ }
\catcode`@=12
\relax
\def\tr{\,{\rm tr}\, }
\def\Im{\,{\rm Im}\, }
\def\Re{\,{\rm Re}\, }

\def\bj{\overline{j}}

\def\bP{\overline{P}}

\def\bE{\overline{E}}
\def\bF{\overline{F}}

\def\bOmega{\overline{\Omega}}
\def\bLambda{\overline{\Lambda}}
\def\thefootnote{\fnsymbol{footnote}}
\def\be{\begin{equation}}
\def\ee{\end{equation}}
\def\ba{\begin{eqnarray}}
\def\ea{\end{eqnarray}}
\def\bs{\begin{subequations}}
\def\es{\end{subequations}}

\def\t{\tau}

\def\gs{g_{\rm string}}

\def\t{\tau}
\def\im{\, {\rm Im}\, \tau}
\def\iT{{\,{\rm Im}\, }T}
\def\iU{{\,{\rm Im}\, }U}
\def\I{\rm I}
\def\II{\rm II}
\def\III{\rm III}

\def\ifd{\int_{\cal F}\frac{d^2\tau}{\im}}

\def\np#1#2#3{Nucl. Phys. {\bf{B#1}} (#2) #3}
\def\pl#1#2#3{Phys. Lett. {\bf{#1B}} (#2) #3}

\def\mpl#1#2#3{Mod. Phys. Lett. {\bf{A#1}} (#2) #3}
\def\nl{\hfil\break}
\def\thebibliography#1{%
\vskip 0.5cm \centerline{\bf References}
\list{%
[\arabic{enumi}]}{\settowidth\labelwidth{[#1]}
\leftmargin\labelwidth
\advance\leftmargin\labelsep
\usecounter{enumi}}
\def\newblock{\hskip .11em plus .33em minus .07em}
\sloppy\clubpenalty4000\widowpenalty4000
\sfcode`\.=1000\relax}

\begin{document}
\renewcommand{\theequation}{\arabic{equation}}
\begin{titlepage}
\begin{flushright}
CERN-TH/96-140 \\
SISSA/81/96/EP \\
LPTENS/96/32 \\
NSF-ITP-96-20 \\
hep-th/9606087 \\
\end{flushright}
\begin{centering}
\vspace{.2in}
{\bf Solving the Decompactification Problem in String Theory}\\
\vspace{0.8 cm}
{E. KIRITSIS$^{\ 1,\, 2}$, C. KOUNNAS$^{\ 1,\, \ast}$,
P.M. PETROPOULOS$^{\ 1,\, \ast}$ \\
\medskip
and \\
\medskip
J. RIZOS$^{\ 1,\, 3,\, \diamond}$}\\
\vskip 0.8cm
{$^1 $\it Theory Division, CERN}\\
{\it 1211 Geneva 23, Switzerland}\\
\medskip
{\it and}\\
\medskip
{$^2 $\it Institute for Theoretical Physics, University of
California}\\
{\it Santa Barbara, CA 93106-4030, USA}\\
\medskip
{\it and}\\
\medskip
{$^3 $\it International School for Advanced Studies, SISSA}\\
{\it Via Beirut 2-4, 34013 Trieste, Italy}\\
\vspace{0.5cm}
{\bf Abstract}\\
\end{centering}
\vspace{.1in}
We investigate heterotic ground states in four dimensions in which
$N=4$ supersymmetry is spontaneously broken to $N=2$. The $N=4$
supersymmetry
is restored at a decompactification limit corresponding to
$m_{3/2}\to 0$. We calculate the full moduli-dependent threshold
corrections and confirm that they are suppressed
in the decompactification limit, as expected from the
restoration of $N=4$ supersymmetry. This should be contrasted with
the
behaviour
of the standard $N=2$ ground states, where the couplings blow up
linearly
with the volume of the decompactifying manifold. This mechanism
provides  a solution to the decompactification problem for
the gauge coupling constants.
 We also discuss how the mechanism can be implemented in
ground states with lower supersymmetry.
\vspace{.1cm}
\begin{flushleft}
CERN-TH/96-140 \\
SISSA/81/96/EP \\
LPTENS/96/32 \\
NSF-ITP-96-20 \\
June 1996
\hfill  To appear in Phys. Lett. {\bf B}\\
\end{flushleft}
\hrule width 6.7cm \vskip.1mm{\small \small \small
$^\ast$\ On leave from {\it Centre National de
la Recherche Scientifique,} France.\\
$^\diamond$\ On leave from {\it Division of Theoretical Physics,
Physics Department,}\\
$^{\ }$\ {\it University of Ioannina,} Greece.}
\end{titlepage}
\newpage
\setcounter{footnote}{0}
\renewcommand{\thefootnote}{\arabic{footnote}}

One of the important unsolved problems in string theory is
supersymmetry breaking, for which there are two known mechanisms.

(\romannumeral1) Tree-level breaking, which comprises Scherk--Schwarz
or internal magnetic type of breaking \cite{ss,b}.
Such an approach can be implemented
in the context of perturbative string theory, but it suffers from the
decompactification problem:
generically, the gravitino mass, which sets the scale of
supersymmetry breaking,
is inversely proportional to an internal radius, and when set to be
of the order of a TeV, a tower of states (charged under low-energy
gauge groups) populate the energy range between the
supersymmetry-breaking
scale and the Planck scale.
Then low-energy couplings run fast (almost linearly with the radius)
and the theory becomes strongly coupled long before the unification
(or Planck) scale.
This behaviour is thought to be generic, and although there are some
ideas on how to avoid it\footnote{The idea consists of constructing
models without $N=2$ sectors, so that the threshold corrections are
independent of the volume moduli of the internal theory.} \cite{ant},
no workable model exists so far.

(\romannumeral2) Non-perturbative breaking via gaugino condensation
\cite{gc}. This is a non-perturbative breaking of supersymmetry,
which up till now had to be discussed at the level of the effective
supergravity. The problem there was the creation of a runaway
potential for the dilaton field, plus our inability to do a
controllable calculation.

In this note we will show that in a class of ground states of
string theory, which have the structure of spontaneously broken $N=4$
theories,
the behaviour of thresholds as functions of the moduli of the
internal
manifold is radically different from the known examples.
The reason for this is that $N=4$
supersymmetry, at large values of (some) moduli,
is restored and thus the thresholds vanish in the limit instead of
blowing up.
Such a behaviour was already anticipated in \cite{kostas}.
Moreover, by using non-perturbative duality techniques, available
today, we might also find a similar behaviour in the context of
non-perturbatively broken
supersymmetry \cite{tobe}.

We will not present here a complete realistic model of broken
supersymmetry.
We expect however that in such models, viewed as orbifolds of $N=4$
ground states
with respect to groups without fixed points (such orbifolds are also
known
as {\it stringy Scherk--Schwarz} ground states \cite{ss}), the
internal
volume dependence of the thresholds comes from $N=2$ sectors.
Thus, here we will study $N=2$ ground states that have the structure
of
spontaneously broken $N=4$ ground states, compute their threshold
corrections
as well as their universal terms, and show that they have the
behaviour
advertised above. A complete analysis, including also $N=1$ and $N=0$
ground states, will appear in \cite{tobe}.

The stringy formula for threshold corrections (for supersymmetric
ground states
and non-anomalous $U(1)$'s) including a stringy infra-red
regularization is
\cite{ka}--\cite{agnt}
\be
{16\, \pi^2\over g_{i}^2(\mu)} = k_{i}{16\, \pi^2\over \gs^2} +
 b_{i}\log {M_{s}^2\over \mu^2} +\Delta_{i}\, ,
\label{1}
\ee
where, in
the $\overline{DR}$ scheme for the effective field theory,
the thresholds read:
\be
\Delta_{i}=\int_{\cal F}{d^2\t\over \im}
\left(
{i\over \pi}
\, {1 \over|\eta|^4}
\sum_{a,b=0,1}
{\partial_{\t}
\vartheta{a\atopwithdelims[]b}\over\eta}
\left(
\bP_i^2 - {k_{i}\over 4\pi{\im}}\right) C{a\atopwithdelims[]b}-
b_{i}
\right)
+b_{i}\log {2\,e^{1-\gamma}\over \pi\sqrt{27}}
\label{2}
\ee
with
\be
b_{i}=
\lim_{\im\to\infty}{i\over \pi}\, {1 \over|\eta|^4}
\sum_{a,b=0,1}
{\partial_{\t}
\vartheta{a\atopwithdelims[]b}\over\eta}
\left(
\bP_i^2 - {k_{i}\over 4\pi{\im}}\right) C{a\atopwithdelims[]b}\, .
\label{2a}
\ee
Here $k_i$ is the level of the $i$th gauge group factor, $b_i$ are
the
full beta-function coefficients, $M_s=1/\sqrt{\alpha'}$ is the
string
scale,
$\mu$ is the infra-red scale, $\bP_i$ is the charge operator of the
gauge
group $G_i$, and
$C{a\atopwithdelims[]b}$ is the internal six-dimensional partition
function.
For conventions see \cite{kk,pr}.

To start with,
we will consider  models
that come from toroidal compactification of generic
six-dimensional
$N=1$ string  theories. In such cases, there is a universal
two-torus, which provides the (perturbative) central charges of the
$N=2$ algebra.
Therefore (\ref{2}) becomes
\be
\Delta_{i}=\int_{\cal F}{d^2\t\over \im}
\Bigg({\Gamma_{2,2}(T,U)\over \bar \eta^{24}}
\left(\bP_{i}^2-{k_{i}\over 4\pi\im}
\right)\overline{\Omega}-b_{i}
\Bigg)+b_{i}\log {2\, e^{1-\gamma}\over \pi\sqrt{27}}\, ,
\label{3}
\ee
where $T$ and $U$ are the complex moduli of the two-torus,
$\overline{\Omega}$ is an antiholomorphic function
and
\ba
\Gamma_{2,2}(T,U)=
\sum_{{\mbox{\footnotesize\bf m}},{\mbox{\footnotesize\bf n}}\in Z}
\exp
\bigg(\!\!\!\!\!\!\!\!\!\!&&
-2\pi i\t\left(
m_{1}\, n_{1}+m_{2}\, n_{2}
\right)\cr &&-
{\pi\im\over \Im T \Im U}
\left|T n_{1} + TUn_{2}+Um_1-m_2\right|^2
\bigg)\, .
\label{4}
\ea

By advocating modular invariance, analytic properties and infra-red
finiteness, it is possible to isolate the universal part of the
thresholds as follows\footnote{Details can be found in \cite{kkpr,
kkpra}.}:
\be
\Delta_{i}=b_{i}^{\vphantom N}\,
\Delta-k_{i}^{\vphantom N} \, Y\, ,
\label{66a}
\ee
where
\be
\Delta=-\log\left(4\pi^2 \big\vert\eta(T)\big\vert^4 \,
\big\vert\eta(U)\big\vert^4
\Im T  \Im U\right)
\label{67}
\ee
and
\be
Y=-{\xi \over 12}\int_{\cal F}{d^2\t\over \im}\,
\Gamma_{2,2}(T,U)\, \Bigg(
\left(\bE_{2}-{3\over \pi\im}\right){\bE_{4}\, \bE_{6}\over \bar
\eta^{24}}-\bj+1008
\Bigg)\, .
\label{68}
\ee
Here $E_{2n}$ are the Eisenstein series:
\be
E_{2}=
{12\over i \pi}\partial_{\t}\log \eta
=1-24\sum_{n=1}^{\infty}{n\, q^n\over 1-q^n}
\label{61b}
\ee
\be
E_{4}=
{1 \over 2}\left(
{\vartheta}_2^8+
{\vartheta}_3^8+
{\vartheta}_4^8
\right)
=1+240\sum_{n=1}^{\infty}{n^3q^n\over 1-q^n}
\ee
\be
E_{6}=
\frac{1}{2}
\left({\vartheta}_2^4 + {\vartheta}_3^4\right)
\left({\vartheta}_3^4 + {\vartheta}_4^4\right)
\left({\vartheta}_4^4 - {\vartheta}_2^4\right)
=1-504\sum_{n=1}^{\infty}{n^5q^n\over 1-q^n}
\, ,\,  \ldots
\ee
and $\xi$ is a constant that can be expressed in terms of the number
of massless vector multiplets $N_V$ and hypermultiplets $N_H$, by
using the relation between gauge and $R^2$-term renormalizations
\cite{kkpr, kkpra}. It reads:
\be
\xi=-{1\over 264}\left(22-N_{V}+N_{H}\right)\, .
\label{70}
\ee
It is remarkable that the latter is {\it fully determined} as a
consequence of the {\it anomaly cancellation} (gauge, gravitational
and mixed) in the underlying six-dimensional theory \cite{sch}.
Indeed, as long as the number of tensor multiplets is $N_T=1$,
anomaly cancellation implies that
\be
N_{H}-N_{V}=242\, ,
\label{20}
\ee
and therefore
\be
\xi=-1\, .
\label{xi1}
\ee

Hence, we observe that for all $N=2$ ground states that come from
toroidal
compactification of a $N=1$ six-dimensional theory, the threshold
corrections
can be completely determined (eqs. (\ref{66a}), (\ref{67}),
(\ref{68}) and (\ref{xi1})) and do not depend on the details of the
six-dimensional theory
(moduli included). As an example, consider
the case of the $Z_{2}$ orbifold, where we have a gauge group
$E_{8}\times
E_7\times SU(2)\times U(1)^2$ and thus $N_V=386$.
The number of massless hypermultiplets is $N_H=628$.
Using these numbers in (\ref{70}) we indeed obtain $\xi=-1$. As
expected from supersymmetry, the corresponding universal threshold is
twice as big as a single-plane contribution of the symmetric
$Z_{2}\times Z_{2}$ orbifold analysed in
\cite{pr}. These thresholds were further analysed in refs.
\cite{kkpr} and \cite{kkpra}. One important feature is that
$Y(T,U)$ given in eq. (\ref{68}) is finite and
continuous inside the moduli space, even along enhanced-symmetry
planes as $T=U$.

We will now construct different $N=2$ models in four dimensions,
which can be
represented as ground states where $N=4$ supersymmetry is
{\it spontaneously} broken to $N=2$.
This can be achieved by doing a $Z_{2}$ rotation on the $T^4$
accompanied by
a $Z_{2}$ translation on the $T^2$.
There are three choices for the $Z_{2}$ translation on the $T^2$:
\nl
(\romannumeral1) $|m_1,m_2,n_1,n_2\rangle$ $\to$ $(-1)^{m_1}$
$|m_1,m_2,n_1,n_2\rangle$,
\nl
(\romannumeral2) $|m_1,m_2,n_1,n_2\rangle$ $\to$ $(-1)^{m_2}$
$|m_1,m_2,n_1,n_2\rangle$,
\nl
(\romannumeral3) $|m_1,m_2,n_1,n_2\rangle$ $\to$ $(-1)^{m_1+m_2}$
$|m_1,m_2,n_1,n_2\rangle$;
\nl
the corresponding models will be refered to as models I, II and III,
respectively.
Here the orbifold group acts without fixed points.
Consequently
the spectrum is in one-to-one correspondence with that of the $N=4$
$T^6$
compactification. This is a spontaneous breaking of $N=4$ to $N=2$
and
the mass of the two gravitinos can be computed:
\nl
(\romannumeral1) $m_{3/2}^2={|U|^2\over \Im T \Im U}$ and
$N=4$ supersymmetry is restored in the limit $\Im T \to \infty$, $U
\to
0$ with
$\Im T \Im U$ finite, and $m^2_{3/2}\to 0$ as it should;
\nl
(\romannumeral2) $m^2_{3/2}={1\over \Im T \Im U}$ and
$N=4$ supersymmetry is restored in the limit $\Im T \to \infty$, $\Im
U \to
\infty$
with $\Im T/\Im U$ fixed;
\nl
(\romannumeral3) $m^2_{3/2}={1\over {\Im T}}\, \inf \left({1\over
\Im U}, {|U|^2\over \Im U} \right)$ and
$N=4$ supersymmetry is restored when $\Im T \to \infty$ with $U$ kept
fixed.

We should note that there also exist models where the lattice
translations
act on the winding numbers. In any such model the original duality
group
$O(2,2;Z)$ is broken to a subgroup. We will describe these groups
below. The broken transformations map any given  model
to a different one. For example the duality transformation
$U\to -1/U$ maps model I to model II.

The heterotic partition functions can be written as
\begin{eqnarray}
Z^A&=&{1 \over \im \vert\eta \vert^4}\,
 {1\over 2}\sum_{a,b=0}^1 (-1)^{a+b+ab}\,
 \left({\vartheta{a\atopwithdelims[]b}\over \eta}\right)^2 \cr
&& {1\over 2}\sum_{h,g=0}^1
 {\vartheta{a+h\atopwithdelims[]b+g}\over \eta}
{\vartheta{a-h\atopwithdelims[]b-g}\over \eta}\,
 Z_{4,4}^{\vphantom A}{h\atopwithdelims[]g}\,
 Z_{2,2}^A {h\atopwithdelims[]g}\cr
&& {1\over 2}\sum_{\bar a,\bar b=0}^1
 {\bar\vartheta{\bar a+h\atopwithdelims[]\bar b+g}\over \bar\eta}
{\bar\vartheta{\bar a-h\atopwithdelims[]\bar b-g}\over \bar\eta}
 \left({\bar\vartheta{\bar a\atopwithdelims[]\bar b\vphantom{g}}\over
 \bar\eta}\right)^6
 {1\over 2}\sum_{\bar c,\bar d=0}^1
 \left({\bar\vartheta{\bar c\atopwithdelims[]\bar d\vphantom{g}}\over
 \bar\eta}\right)^8\, ;
\label{22}
\end{eqnarray}
here the index $A$ labels the three possible translations I, II, III,
and
$$
Z_{4,4}{0\atopwithdelims[]0}
={\Gamma_{4,4}\over
|\eta|^8}
\ , \ \
Z_{4,4}{0\atopwithdelims[]1}
=16\, {|\eta|^4\over
|\vartheta_2|^4}={|\vartheta_3\, \vartheta_4|^4\over |\eta|^8}\ ,
$$
\be
Z_{4,4}{1\atopwithdelims[]0}
=16\, {|\eta|^4\over
|\vartheta_4|^4}={|\vartheta_2\, \vartheta_3|^4\over
|\eta|^8}
\ , \ \
Z_{4,4}{1\atopwithdelims[]1}
=16\, {|\eta|^4\over
|\vartheta_3|^4}={|\vartheta_2\, \vartheta_4|^4\over |\eta|^8}\ .
\label{24}
\ee
The shifted partition functions of the two-torus
are
\be
Z_{2,2}^A{h\atopwithdelims[]g}(T,U)
={\Gamma_{2,2}^A{h\atopwithdelims[]g}(T,U)\over
|\eta|^4}\, ,
\label{25}
\ee
where $\Gamma_{2,2}^A{0\atopwithdelims[]0}
\equiv \Gamma_{2,2}^{\vphantom A}$ is given in (\ref{4}), and
\nl
(\romannumeral1)
$\Gamma^{\I}_{2,2}{h\atopwithdelims[]g}$ is obtained from
$\Gamma_{2,2}^{\vphantom A}$ by inserting
$(-1)^{m_1 \, g}$ and shifting $n_1\to n_1+h/2$,
\nl
(\romannumeral2)
$\Gamma^{\II}_{2,2}{h\atopwithdelims[]g}$
is obtained from
$\Gamma_{2,2}^{\vphantom A}$ by inserting
$(-1)^{m_2 \, g}$ and shifting $n_2\to n_2+h/2$,
\nl
(\romannumeral3)
$\Gamma^{\III}_{2,2}{h\atopwithdelims[]g}$
is obtained from
$\Gamma_{2,2}^{\vphantom A}$ by inserting
$(-1)^{\left(m_1+m_2\right) g}$ and shifting $n_1\to n_1+h/2$ and
$n_2\to
n_2+h/2$.
These partition functions can be obtained from
\ba
\Gamma_{2,2}{h_1,h_2\atopwithdelims[]g_1,g_2}=
\sum_{{\mbox{\footnotesize\bf m}},{\mbox{\footnotesize\bf n}}\in Z}
(-1)^{m_1 \, g_1+m_2 \, g_2}
\exp
\Bigg(-
2\pi i\t\left(
m_{1}\left( n_{1}+{h_1 \over 2}\right)+m_{2}\left( n_{2}+{h_2 \over
2}\right)
\right)\cr -
{\pi\im\over \Im T \Im U}
\left|T \left( n_{1}+{h_1 \over 2}\right) + TU\left( n_{2}+{h_2 \over
2}\right)+Um_1-m_2\right|^2
\Bigg)\, ,
\label{44444}
\ea
where
${h_1,h_2\atopwithdelims[]g_1,g_2}$ takes the following values:
${h,0\atopwithdelims[]g,0}$,
${0,h\atopwithdelims[]0,g}$ and
${h,h\atopwithdelims[]g,g}$ for models
I, II and III, respectively. We also have the periodicity
\be
\Gamma^{A}_{2,2}{h\atopwithdelims[]g}=
\Gamma^{A}_{2,2}{h+2\atopwithdelims[]g}=
\Gamma^{A}_{2,2}{h\atopwithdelims[]g+2}\, ,
\label{25a}
\ee
and the modular properties
\be
\tau\to\tau+1 \ ,\ \ Z^{A}_{2,2}{h\atopwithdelims[]g}\to
Z^{A}_{2,2}{h\atopwithdelims[]h+g}
\label{255}\ee
\be
\tau\to-{1\over \tau} \ ,\ \ Z^{A}_{2,2}{h\atopwithdelims[]g}\to
Z^{A}_{2,2}{g\atopwithdelims[]h}\, .
\label{256}\ee
Finally,
by straightforward computation we obtain:
\be
{1\over 2}
\sum_{h,g=0}^1\Gamma^{\I}_{2,2}{h\atopwithdelims[]g}(T,U)=
\Gamma_{2,2}^{\vphantom A}\left({T\over 2},2U\right)
\label{26a}
\ee
\be
{1\over 2}
\sum_{h,g=0}^1\Gamma^{\II}_{2,2}{h\atopwithdelims[]g}(T,U)=
\Gamma_{2,2}^{\vphantom A}\left({T\over 2},{U\over 2}\right)
\label{26b}
\ee
\be
{1\over 2}
\sum_{h,g=0}^1\Gamma^{\III}_{2,2}{h\atopwithdelims[]g}(T,U)=
\Gamma_{2,2}^{\vphantom A}\left({T\over 2},{1+U\over 1-U}\right)\, ,
\label{26c}
\ee
which will be useful further on.

The above models have the same gauge group as that of the standard
$Z_2$
orbifold limit of $K_3$ so that $N_V=386$.
However the number of massless hypermultiplets is different. Compared
with the usual $Z_2$ orbifold model, the massless hypermultiplets
coming from the twisted sector have become massive here. Thus the
massless ones are 4 singlets
under $E_8\times E_7\times SU(2)$ and one which is singlet under
$E_8$ and transforms under $E_7\times SU(2)$ as $(56,2)$;
then $N_H=116$.
This should be compared with the standard $Z_2$ orbifold, which
contains
extra massless hypermultiplets coming from the twisted sector: 8
transforming as $(56,1)$ and $32$ as $(1,2)$ under $E_7\times SU(2)$.

A comment is also in order here concerning the duality symmetries of
these models.
The standard $Z_2$ orbifold has an $O(2,2;Z)$ duality symmetry
acting on the $T,U$ moduli, which is generated by the standard
$SL(2;Z)$ transformations on both $T$ and $U$ plus the
$T\leftrightarrow U$
interchange transformation.
The duality group in this case can be found from the explicit form
of the
$O(2,2;Z)$ transformations:
\be
SL(2;Z)_T \ :\ \
\left(\matrix{m_1 \cr m_2 \cr n_1 \cr n_2 \cr}\right)
\to
\left(
\matrix{
d&{\hphantom{-}}0&0&b\cr 0&{\hphantom{-}}d&-b{\hphantom{-}}&0\cr
0&-c&a&0\cr c&{\hphantom{-}}0&0&a\cr}
\right)
\left(
\matrix{m_1 \cr m_2 \cr  n_1 \cr n_2\cr}
\right)\ ,\ \ ad-bc=1\, ,
\label{261}\ee
\be
SL(2;Z)_U \ :\ \
\left(\matrix{m_1 \cr m_2 \cr n_1 \cr n_2 \cr}\right)
\to
\left(
\matrix{
{\hphantom{-}}a'&-c'{\hphantom{-}}&0&0\cr -b'&d'&0&0\cr
{\hphantom{-}}0&0&d'&b'\cr {\hphantom{-}}0&0&c'&a'\cr}
\right)
\left(
\matrix{m_1 \cr m_2 \cr  n_1 \cr n_2\cr}
\right)\ ,\ \ a'd'-b'c'=1
\label{262}\ee
and
\be
T\leftrightarrow U \ :\ \
\left(
\matrix{m_1 \cr m_2 \cr n_1 \cr n_2 \cr}
\right)
\to
\left(
\matrix{0&0&1&0\cr 0&1&0&0\cr 1&0&0&0\cr 0&0&0&1\cr}
\right)
\left(\matrix{m_1 \cr m_2 \cr n_1 \cr n_2 \cr}\right)\, .
\label{263}
\ee
The above transformations can be summarized as follows:
\be
\left(\matrix{{\bf m}\cr {\bf n}\cr}\right)
\to
\left(
\matrix{
A & B \cr C & D \cr}\right)
\left(\matrix{{\bf m} \cr {\bf n} \cr}
\right)
\ ,\ \
A,B,C,D\in GL(2;Z)
\label{264}\ee
with
\be
C^T\,A+A^T\,C=0\ ,\ \ D^T\,B+B^T\,D=0\ ,\ \
C^T\,B+A^T\,D=1\, .
\label{265}\ee
(\romannumeral1) For model I the duality group is $\Gamma(2)_T$,
defined by
$b$ even, times $\Gamma(2)_U$, defined by $c'$ even.
In the notation of (\ref{264}) this corresponds to
$A_{12}, B_{12}, B_{21}, D_{21}$
being even and $B_{11}=0\ {\rm mod}(4)$.\nl
(\romannumeral2) For model II the duality group is $\Gamma(2)_T$,
defined by
$b$ even, times $\Gamma(2)_U$, defined by $b'$ even as well as the
$T\leftrightarrow U$ interchange.
In the notation of (\ref{264}) this corresponds to
$A_{21}, B_{12}, B_{21}, D_{12}$
being even and $B_{22}=0\ {\rm mod}(4)$.\nl
(\romannumeral3) For model III the duality group is
$\Gamma(2)_T$ defined by $b$ even, times $\tilde\Gamma_U$.
The group $\tilde\Gamma$ is defined by the integer matrices
\be
\left(\matrix{m&2n+m+1\cr 2r-m+1& 2s-m\cr}\right)
\ee
with determinant one.
In the notation of (\ref{264}) this corresponds to
$A_{11}+A_{12}$ and $A_{21}+A_{22}$
being simultaneously even or simultaneously
odd, $D_{11}+D_{12}$ and $D_{21}+D_{22}$
being simultaneously even or simultaneously
odd, and
$B_{11}+B_{12}+B_{21}+B_{22}$
vanishing modulo $4$ together with the other combinations where
two signs are flipped.

We will now evaluate the threshold corrections for the above three
models.
Using eq. (\ref{2}) we obtain\footnote{The prime summation
over $(h,g)$ stands for $(h,g) =
\{(0,1)$,
$(1,0)$, $(1,1)\} $.}:
\be
\Delta^{A}_{i}=\int_{\cal F}{d^2\t\over \im}
\left(-\sump{\Gamma_{2,2}^A\oao{h}{g}\over \bar \eta^{24}}
\left(\bP_{i}^2-{k_{i}\over 4\pi\im}
\right)\overline{\Omega}\oao{h}{g}-b_{i}^{\vphantom A}
\right)+b_{i}^{\vphantom A}\log {2\, e^{1-\gamma}\over \pi\sqrt{27}}
\label{27}
\ee
with
\ba
\Omega\oao{0}{1}&=&{\hphantom{-}}
{1\over 2}\, E_4\, \vartheta^4_3\, \vartheta_4^4
\left(\vartheta^4_3+\vartheta_4^4\right)\label{28a}\\
\Omega\oao{1}{0}&=&-
{1\over 2}\, E_4\,\vartheta^4_2\, \vartheta_3^4
\left(\vartheta^4_2+\vartheta_3^4\right)\label{28b}\\
\Omega\oao{1}{1}&=&-
{1\over 2}\, E_4\,\vartheta^4_2\, \vartheta_4^4
\left(\vartheta^4_2-\vartheta_4^4\right)\, .\label{28c}
\ea
These functions obey the following identity:
\be
\sump\Omega\oao{h}{g}=E_4\,  E_6\label{29}\, .
\ee
We proceed by  using arguments similar to those that were used
in refs. \cite{kkpr, kkpra} in order to reach (\ref{68}) starting
from
(\ref{3}).
The thresholds (\ref{27}) can be written in the form
\be
\Delta^A_i= \ifd \left(\bF^A_i-b_i^{\vphantom A}\right)+
b_{i}^{\vphantom A}\log {2\, e^{1-\gamma}\over
\pi\sqrt{27}}
\label{31}
\ee
with
\ba
\bF^{A}_{i} &=&- k_i^{\vphantom A} \sump \frac{\Gamma^{A}_{2,2}
\oao{h}{g}}{\bar{\eta}^{24}}
\left(\frac{\bE_2}{12} -
\frac{1}{4 \pi
\im}\right)\bOmega\oao{h}{g}
-  \sump\frac{\Gamma^{A}_{2,2}
\oao{h}{g}}{\bar{\eta}^{24}}\left(\bP^2_i -
\frac{k_i\, \bE_2}{12}\right)\bOmega
\oao{h}{g}
\cr
&=&{\hphantom -}k_i^{\vphantom A}\, \bF^A_{{\rm grav}}  +
 \sump\Gamma^{A}_{2,2}
\oao{h}{g}\, \bLambda_i{\vphantom A}\oao{h}{g}\, .
\ea
Here we have separated the universal term
\be
 \bF^A_{{\rm grav}} =  -\sump \frac{\Gamma^{A}_{2,2}
\oao{h}{g}}{\bar{\eta}^{24}}
\left(\frac{\bE_2}{12} -
\frac{1}{4 \pi
\im}\right)\bOmega\oao{h}{g}
\, ,
\ee
which is the associated function that appears in the $R^2$-term
renormalization,
and
\be
\bLambda_i\oao{h}{g} = -\frac{1}{\bar{\eta}^{24}}\left(\bP^2_i -
\frac{k_i\, \bE_2}{12}\right)\bOmega
\oao{h}{g}
\, .
\ee
For the models under consideration, $\Lambda_i\oao{h}{g}$ can be
expressed as
$\Lambda_i\oao{0}{1} = f_i(1-x)$, $\Lambda_i\oao{1}{0} = f_i(x)$,
$\Lambda_i\oao{1}{1} = f_i\left(\frac{x}{x-1}\right)$, where
$x=\left(\vartheta_2 / \vartheta_3\right)^4$
and
\be
f_i(x) =  b_i - k_i  \, \rho(x) +
 \left(\frac{\tilde b_i}{3}-b_i - 40\, k_i\right)\sigma(x)
\label{eeaa}
\ee
with
\ba
\rho(x) &=& {\hphantom -}\frac{4}{3\, x(x-1)^2}\left(8-49\, x+66\,
x^2-49\, x^3+8 \, x^4\right)
\cr
\sigma(x) &=&-\frac{(x-1)^2}{x}\, .
\ea
The constants $ b_i$ are the beta-function coefficients of the models
($b_{E_8} = -60$, $b_{E_7}=-12$, $b_{SU(2)}=52$) while
 ${\tilde b_i}$ are the beta-function coefficients of the symmetric
$Z_2$ orbifold
($\tilde{b}_{E_8} = -60$, $\tilde{b}_{E_7}=\tilde{b}_{SU(2)}=84$).
Note also the identities\footnote{We use the notation
$\sigma\oao{0}{1}=\sigma(1-x)$,
$\sigma\oao{1}{0}=\sigma(x)$, $\sigma\oao{1}{1}=
\sigma\left({x \over x-1}\right)$, and
similarly
for $\rho(x)$.}
\ba
\sump\sigma\oao{h}{g} &=& 3\cr
\sump\rho\oao{h}{g} &=& -\frac{j}{12} -36\, ,
\ea
which imply
\be
\sump f_i\oao{h}{g} =  k_i\left(\frac{j}{12} -84\right)
+\tilde{b}_i\, ,
\ee
and lead to  the correct result (\ref{67}), (\ref{68})
for the
symmetric $Z_2$
orbifold obtained when the substitution
$\Gamma^{A}_{2,2}\oao{h}{g} \to
\Gamma_{2,2}^{\vphantom A}$ is performed.

Using  the above results, the thresholds (\ref{31}) take the form
\be
\Delta^A_i = b_i^{\vphantom A} \, \Delta^A_{\vphantom i} +
\left(\frac{\tilde{b}_i}{3} -b_i^{\vphantom
A}\right)\delta^A_{\vphantom i}
-k_i^{\vphantom A}\,  Y^A_{\vphantom i}\, ,
\ee
where
\ba
\Delta^A &=& \ifd\left(\sump\Gamma^{A}_{2,2}\oao{h}{g}-1\right)
+ \log {2\, e^{1-\gamma}\over \pi\sqrt{27}}
\label{32}\cr
\delta^A &=& \ifd\sump\Gamma^{A}_{2,2}\oao{h}{g}\,
\bar{\sigma}\oao{h}{g}\label{32a}\cr
Y^A &=& \ifd\sump\Gamma^{A}_{2,2}\oao{h}{g}\left(
\frac{1}{\bar{\eta}^{24}}
\left(\frac{\bE_2}{12} -
\frac{1}{4 \pi
\im}\right)\bOmega\oao{h}{g}
+\bar{\rho}\oao{h}{g} + 40\,\bar{\sigma}\oao{h}{g}\right)\, .
\label{32b}
\ea

Before any further computation we should address two issues here.
First we observe that the threshold corrections in these models
cannot be decomposed in the simple form $b_i \, \Delta +k_i \, Y$, as
was described
earlier
for $N=2$ models
coming from toroidal compactification of six-dimensional  theories.
The other issue is why the beta-function coefficients $\tilde b_i$ of
the
previous models come in the thresholds here.
This is not difficult to understand, once we consider some
decompactification limits of the present models.
The first is the supersymmetry restoration limit $m_{3/2}\to 0$.
In this case we obtain a $N=4$ theory in five or six dimensions.
There is, however, another decompactification limit, namely
$m_{3/2}\to
\infty$.
There we obtain a six-dimensional $N=1$ theory, which is exactly the
same as the one relevant for the original models.
The above statements come from the following decompactification
limits of the two-torus blocks.
Consider (\ref{44444}) in the Lagrangian representation:
\be
\!\!\!
\im \, \Gamma_{2,2}{h_1,h_2\atopwithdelims[]g_1,g_2}=\sqrt{\det G}
\sum_{{\mbox{\footnotesize\bf m}},{\mbox{\footnotesize\bf n}}\in
Z}e^{-{\pi\over
\im}
\sum_{i,j}
\left(G_{ij}+B_{ij}\right)\left[m_{i}+{g_i\over 2}+
\left(n_i+{h_{i}\over
2}\right)\tau\right] \left[m_{j}+{g_j\over 2}+
\left(n_j+{h_{j}\over 2}\right)\bar \tau\right]}\, ,
\label{I1}
\ee
where as usual
\be
G={\iT \over \iU}
\left(\matrix{1 & \Re U \cr \Re U & |U|^2 \cr}\right)
\ ,\ \ B=\Re T \left(\matrix{ 0 & -1 \cr
1 & {\hphantom -}0 \cr}\right)\, .
\label{I2}\ee
It is easy to verify that as $\iT \to \infty$:
\be
\im \, \Gamma_{2,2}{h_1,h_2\atopwithdelims[]g_1,g_2} \to
\left\{ \begin{array}{l}
\iT
\ {\rm for}\ h_i,g_i=0\, ,\cr
{\rm exponentially \  suppressed \ otherwise}\, .\end{array}
\right.
\label{I3}
\ee
Using the dual partition function we can check that in the opposite
limit,
$\iT \to 0$:
\be
\im \, \Gamma_{2,2}{h_1,h_2\atopwithdelims[]g_1,g_2} \to
{1\over \iT}
\  \ \forall \ h_i,g_i\, .
\label{I4}\ee
However, from the six-dimensional view point,
the beta-function coefficients $\tilde
b_i$ are related to anomaly
coefficients in that theory (see for example \cite{AFIQ}).
Consider the anomaly eight-form in six dimensions $I_8=X_4\, \tilde
X_4$
with
\be
X_4={1\over 4(2\pi)^2}
\left(\tr R^2_{\vphantom a}-\sum_i v_i^{\vphantom 2}\tr F^2_i\right)
\ ,\ \
\tilde X_4={1\over 4(2\pi)^2}
\left(\tr R^2_{\vphantom a}-\sum_i \tilde v_i^{\vphantom 2} \tr
F^2_i\right)\, .
\label{I5}
\ee
The first term appears in the six-dimensional Bianchi identity, $d
H=\alpha'
(2\pi)^2 X_4$
while the second is the one-loop correction of the field equation
$d \star H=\alpha' (2\pi)^2 \tilde X_4$.
The coefficients $v_i$ are related to the tree-level value of the
gauge couplings, namely $v_i=k_i/c_i$ with $c_i=2,1,{1\over
3},{1\over 6},
{1\over 30}$ for
$SU(N)$, $SO(2N)$,
$E_{6,7,8}$.
The coefficients $\tilde v_i$ are determined from Green--Schwarz
anomaly cancellation, and the
corresponding four-dimensional
$N=2$ beta-function coefficients are
given by
\be
\tilde b_i=12\left(1+{\tilde v_i\over v_i}\right)\, .
\label{I6}
\ee
This explains their appearance in our models.

The integrals in (\ref{32}) can be explicitly evaluated using the
results of \cite{dkl,hm}, (\ref{26a})--(\ref{26c}) and the
duplication
formulae for the
$\vartheta$-functions with the result\footnote{Similar non-universal
thresholds in the context of $N=1$ models have also been computed
(and presented in a different form) in \cite{sabra}.}:
\be
\Delta^{\I}=-\log\left({\pi^2\over 4}
\,
\big|\vartheta_4(T) \big|^4\,
\big|\vartheta_2(U) \big|^4
\iT \iU \right)
\label{34a}
\ee
\be
\Delta^{\II}=-\log\left({\pi^2\over 4}
\,
\big|\vartheta_4(T) \big|^4\,
\big|\vartheta_4(U) \big|^4
\iT \iU \right)
\label{34b}
\ee
\be
\Delta^{\III}=-\log\left({\pi^2\over 4}
\,
\big|\vartheta_4(T) \big|^4\,
\big|\vartheta_3(U) \big|^4
\iT \iU \right)\, .
\label{34c}
\ee
These thresholds should be contrasted with (\ref{67}).
For both decompactification limits $\iT\to\infty$ and $\iU\to \infty$
in (\ref{67}),
we obtain the familiar result that the thresholds grow linearly with
the appropriate modulus.
To simplify matters we will take $\Re T=\Re U=0$ so that
$\iT= R_1 \, R_2$, $\iU=R_2/R_1$, and the two-torus decouples into
two
circles
with radii $R_1$ (associated with $m_1,n_1$) and $R_2$ (associated
with $m_2,n_2$), respectively.
The thresholds for the original models take the form ($R_1 =
R_2 = R$)
\be
\lim_{ R\to\infty} \Delta ={\pi\over 3}\, R^2-\log R^2+
{\cal O}(1)\, .
\label{35}\ee
This is the expected {\it geometric} behaviour of the running gauge
coupling.
Similarly we also have, for the universal contribution \cite{pr}:
\be
\lim_{R\to\infty} Y=4\pi R^2 +{\cal O}\left({1\over R^2}\right)\, .
\label{36}\ee
Notice the absence of the logarithmic piece in (\ref{36}).
We will also consider for
the sake of comparison the limit $R_1\to \infty$ for fixed $R_2$.
We obtain:
\be
\lim_{ R_1\to\infty} \Delta={\pi\over 3}\left(R_2^{\vphantom
2}+{1\over
R_2}\right) R_1^{\vphantom 2}
-\log R_1^2+{\cal O}(1)
\label{I66}\ee
while  for the universal term
\be
\lim_{R_1\to\infty} Y={4\pi\over 3}\,  R_1
\left(3 R_2 +
\frac{1}{R_2^3}\right)\Theta(R_2-1)
+{4\pi\over 3}\, R_1
\left({3\over  R_2}
+R_2^3\right)\Theta(1-R_2)+
{\cal O}\left({1\over R^2_1}\right)\, .
\label{I67}\ee

Let us now compute the large-radius limits of the thresholds
(\ref{32b}).
For model I, the supersymmetry restoration
 limit is $R_1\to \infty$ ($m_{3/2}=1/R_1$). In this limit we obtain:
\ba
\lim_{R_{1}\to\infty}\Delta^{\I}&=&-\log\left(R_1^2\right)+
{\cal O}(1)\cr
\lim_{R_{1}\to\infty}\delta^{\I} &=& {\cal
O}\left(e^{-R_1^2}\right)\, .
\label{37}
\ea
As expected we no longer have the linear explosion of the
threshold correction as in (\ref{I66}), but we obtain the
(apparently)
counter-intuitive result
that the threshold correction diverges logarithmically. Indeed,
we would expect that since in this limit $N=4$ supersymmetry is
restored,
the thresholds should vanish.
Below we will resolve this discrepancy, which is an infra-red
phenomenon.
Notice, however, that since  the universal contribution is
infra-red-finite,
it vanishes in the restoration limit, as it should:
\be
\lim_{R_{1}\to\infty}Y^{\I}={\cal O}\left({1\over R_1^2}\right)\, .
\label{38}\ee
The asymptotic expressions are similar for model II (with
$R_1\leftrightarrow
R_2$) while for model III, where both $R_1\to\infty$ and
$R_2\to\infty$,
we
get
\ba
\lim_{R_{1,2}\to\infty}\Delta^{\III}&=&-\log\left({ R_1 \,
R_2}\right)+
{\cal O}(1)\cr
\lim_{R_{1,2}\to\infty}\delta^{\III} &=& {\cal O}\left(e^{-R_1\,
R_2}\right)\\
\lim_{R_{1,2}\to\infty}Y^{\III}&=&{\cal O}\left({1\over R_1\,
R_2}\right)\, .
\label{37aa}
\ea
Note also that in the opposite limit $m_{3/2}\to \infty$ the
thresholds
match those (\ref{35})--(\ref{I67}) of the previously described
models
\cite{kkpr, kkpra}.

Let us now concentrate on
the issue of how to take various infra-red limits.
Consider turning on all Wilson lines which break the gauge group to
$U(1)$'s.
All charged states are now massive.
Some of these states in general have masses of the same order
as $m_{3/2}$,
or
above, and some others have masses below $m_{3/2}$.
Suppose first that there are no states with masses lower than
$m_{3/2}$.
In this case, as $m_{3/2}\to 0$, all thresholds would vanish in the
limit.
In the case where there are states that are always lighter than
$m_{3/2}$
as $m_{3/2}\to 0$, these states
have always an effective $N=2$ behaviour and they will produce a
logarithmic
running even when $m_{3/2}=0$.
In the case we considered, the charged massless states produce an
infra-red
divergence proportional to $\log\mu^2$, which, when present, implies
also
the (non-holomorphic) logarithmic factor $\log(\iT \iU)$.

The question we should answer is: Can we turn Wilson lines so that
all charged states become heavier than $m_{3/2}$?
The answer is yes.
Consider for simplicity model I, where the gravitino mass is (in the
presence of Wilson lines $W_i^{\vphantom 1}=Y^1_i+U\, Y^2_i$):
\be
m^2_{3/2}={|U|^2\over \iT \iU-\sum_i \Im W_i^2}\, .
\label{I36}\ee
The masses of charged particles, which become massless for zero
Wilson
lines, are
\be
m^2_{\rm charged}={\big|W^i q_i \big|^2 \over \iT \iU-
\sum_i\Im W_i^2}\, .
\label{I37}\ee
where $q_i$ are integers.
Due to duality symmetries, the Wilson lines have the periodicity
properties $W_i\sim W_i+1\sim W_i+U$.
It is now obvious that for finite $Y_i^1$, in the supersymmetry
restoration
limit $U\to 0$, all charged particles become heavier than the
gravitinos.
Similar remarks apply to the other models.

For practical purposes, the supersymmetry-breaking scale is a tiny
fraction
of the Planck mass and there are few states that lie below it. Thus,
even in such a case there is no decompactification problem. There is
only the logarithmic running of the light particles up to the
supersymmetry-breaking scale. Anything
above does not run since the physics above $m_{3/2}$ is governed by
$N=4$ supersymmetry.

Can this mechanism work in models with lower supersymmetry?
In the most general case we could imagine three ordered
supersymmetry-breaking scales, $m_0\leq m_1\leq m_2$. At scales above
$m_2$ we have
restored $N=4$ supersymmetry. In the range $m_1\leq \mu\leq m_2$
supersymmetry is spontaneously broken to $N=2$ while for $m_0\leq
\mu\leq m_1$ it is spontaneously broken to $N=1$. Finally,
below $m_0$ all supersymmetries are
broken.
It is possible that some of these scales are the same, e.g.
$m_0=m_1$.
If $m_2$ is sufficiently low then there is no decompactification
problem
in the theory. It remains to be seen  if realistic models can be
constructed
with low values of $m_2$.
\vskip 0.3cm
\centerline{\bf Acknowledgements}

E. Kiritsis was supported in part by the grant PHY94-07194.
He would also like to thank the ITP for the warm hospitality extended
to him
during his stay there, while part of this work was done.
C. Kounnas was  supported in part by EEC contracts
SC1$^*$-0394C and SC1$^*$-CT92-0789.
J. Rizos would like to thank the CERN Theory Division  for
hospitality
and  acknowledges financial
support from the EEC contract {ERBCHBGCT940634}.
\eject


\begin{thebibliography}{99}

\bibitem{ss}
C.~Kounnas and M.~Porrati, \np {310} {1988} {355};\\
S.~Ferrara, C.~Kounnas, M.~Porrati and F.~Zwirner,
\np {318} {1989} {75};\\
I. Antoniadis, C. Bachas, D. Lewellen and T. Tomaras, \pl
{\bf 207} {1988} {441};\\
C.~Kounnas and B.~Rostand, \np {341} {1990} {641};\\
I.~Antoniadis, \pl {246} {1990} {377};\\
I.~Antoniadis and C.~Kounnas, \pl {261} {1991} {369}.

\bibitem{b} C. Bachas, hep-th/9503030.

\bibitem{ant} I. Antoniadis, \pl {246} {1990} {377};\\
K. Benakli, hep-th/9509115.

\bibitem{gc}
H.-P.~Nilles, \pl {115} {1982} {193} and \np {217} {1983}
{366};\\
S.~Ferrara, L.~Girardello and H.-P.~Nilles, \pl {125} {1983}
{457};\\
J.-P.~Derendinger, L.E.~Ib\'a\~nez and H.-P.~Nilles, \pl {155}
{1985} {65};\\
M.~Dine, R.~Rohm, N.~Seiberg and E.~Witten, \pl {156} {1985}
{55};\\
C.~Kounnas and M.~Porrati, \pl {191} {1987} {91}.

\bibitem{kostas} S.~Ferrara, C.~Kounnas and F.~Zwirner,
\np{429} {1994} {589} and erratum, ibid. {\bf B433} (1995)
255.

\bibitem{tobe} E. Kiritsis and C. Kounnas, CERN-TH/96-134;\\
E. Kiritsis, C. Kounnas, P.M. Petropoulos and J. Rizos, in
preparation.

\bibitem{ka} V. Kaplunovsky, \np{307}{1988}{145} and erratum,
ibid. {\bf B382} (1992) 436.

\bibitem{dkl} L. Dixon, V. Kaplunovsky and J. Louis,
\np{355}{1991}{649}.

\bibitem{rest} J.-P.~Derendinger, S.~Ferrara, C.~Kounnas and
F.~Zwirner, \np {372} {1992} {145} and \pl {271} {1991} {307};\\
S. Ferrara, C. Kounnas, D. L\"ust and F. Zwirner, \np {365}
{1991} {431};\\
I. Antoniadis, K.S. Narain and T. Taylor, \pl {267}
{1991} {37};\\
G.~Lopes Cardoso and B.A.~Ovrut, \np {369} {1992}{351};\\
I. Antoniadis, E. Gava, K.S. Narain and T. Taylor, \np {407}
{1993} {706} and ibid. {\bf B413}  (1994) 162;\\
P. Mayr and S. Stieberger, \np{407}{1993}{725};\\
D. Bailin, A. Love, W.A. Sabra and S. Thomas,
\mpl{9}{1994}{67};\\
V. Kaplunovsky and J. Louis, \np{444}{1995}{191}.

\bibitem{kk} E. Kiritsis and C. Kounnas, Nucl. Phys. {\bf B41}
[Proc. Sup.] (1995) 331; \np {442} {1995} {472};
proceedings of {\sl STRINGS 95, Future Perspectives in String
Theory,}
Los Angeles, CA, 13--18 March 1995, hep-th/9507051.

\bibitem{sabra} D. Bailin, A. Love, W.A. Sabra and S. Thomas,
\mpl{10}{1995}{1337}.

\bibitem{hm} J.A. Harvey and G. Moore, \np{463}{1996}{315}.

\bibitem{pr} P.M. Petropoulos and J. Rizos,  \pl{374}{1996}{49}.

\bibitem{kkpr}  E. Kiritsis, C. Kounnas, P.M. Petropoulos and J.
Rizos,
 hep-th/9605011.

\bibitem{kkpra}  E. Kiritsis, C. Kounnas, P.M. Petropoulos and J.
Rizos,
hep-th/9608034.

\bibitem{agnt} I. Antoniadis, E. Gava and K.S. Narain, \pl {283}
{1992}
{209} and
\np {383} {1992} {93}.

\bibitem{sch} J.H. Schwarz, \pl{371}{1996}{223}.

\bibitem{AFIQ} G. Aldazabal, A. Font, L.E. Ib\'an\~ez and F. Quevedo,
hep-th/9602097.

\end{thebibliography}
\end{document}